\documentclass[12pt, letterpaper]{article}
 
\usepackage{graphicx} 
\usepackage{color}
\usepackage{setspace} 
\usepackage[longnamesfirst]{natbib} 
\usepackage{booktabs} 
\usepackage{rotating} 
\usepackage{amsmath} 
\usepackage{fullpage} 
\usepackage{subfig}
\usepackage{mdwlist}
\usepackage{authblk}
\usepackage{url}
\usepackage{multirow}
\usepackage[nolists]{endfloat} 
\bibpunct{(}{)}{;}{a}{}{,} 
\newcommand{\cption}[1]{\caption{\doublespacing #1}} 

\title{The Role of Race, Ethnicity, and Gender in the Congressional Cosponsorship Network}

  \author[1]{Alison Craig \thanks{craig.373@buckeyemail.osu.edu}}
    \author[1]{Skyler J. Cranmer \thanks{cranmer.12@osu.edu}}
 \author[2]{Bruce A. Desmarais\thanks{bdesmarais@psu.edu}}
   \author[3]{Christopher J. Clark \thanks{chriclar@email.unc.edu}}
 \author[4]{Vincent G. Moscardelli \thanks{vin.moscardelli@uconn.edu}}
 \affil[1]{\small{Department of Political Science, Ohio State University}}
 \affil[2]{\small{Department of Political Science, Pennsylvania State University}}
  \affil[3]{\small{Department of Political Science, University of North Carolina at Chapel Hill}}
 \affil[4]{\small{Department of Political Science, University of Connecticut}}

\begin{document} 
\maketitle
\vspace{-.4cm}

\begin{abstract}
\noindent Previous research indicates that race, ethnicity, and gender influence legislative behavior in important ways.  The bulk of this research, however, focuses on the way these characteristics shape an individual legislator's behavior, making it less clear how they account for relationships between legislators. We study the cosponsorship process in order to understand the race and gender based dynamics underlying the relational component of representation.  Using a temporal exponential random graph model, we examine the U.S. House cosponsorship network from 1981 through 2004.  We find that Black and Latino members of Congress are at a comparative disadvantage as a result of race-based assortative mixing in the cosponsorship process, yet this disadvantage is mitigated by the electoral pressures that all members face.  Members representing districts with significant racial and ethnic minority populations are more likely to support their minority colleagues.  We also find that women members do not appear to face a similar disadvantage as a result of their minority status.  We argue that these race and gender dynamics in the cosponsorship network are the result of both the inherent tendency towards intra-group homophily in social networks and the electoral connection, which is manifested here as members supporting minority colleagues to broaden their own electoral base of support among minority constituencies.
\end{abstract}
\clearpage
\doublespacing
The 114th Congress was heralded as the most diverse Congress in history, yet it remains disproportionately white and male when compared to the population of the United States.  Of the 535 members of Congress who took office in 2015, 104 were women and 96 were racial or ethnic minorities.\footnote{While this is a marked increase from the 56 who were women and 68 who were racial or ethnic minorities twenty years ago, women and minorities remain significantly underrepresented in the United States Congress.  Blacks comprise 13\% of the US population, but only 8.5\% of Congress.  Latinos comprise 17\% of the population, but only 7\% of Congress.  The disparity is largest for women, who make up 51\% of the population, but only 20\% of the representatives in Congress.} The disproportionately white and male membership of the United States Congress has important implications for how women and minorities are represented, as has been described by a multitude of scholars \citep[e.g.][]{Baker:2005, Canon:1999, Hero:1995, Mansbridge:1999, Tate:2003}.  However, in addition to being a problem for descriptive representation, this underrepresentation of minorities in Congress has potential implications for how effective they can be as legislators. For a legislator to be effective, collaboration and coalition-building are key. Members of Congress must rely on others to advance their agendas, from gathering cosponsors to ensuring a bill has enough votes to pass.  Building  a broad coalition usually means reaching across the aisle in the search for collaborators \citep{Fenno:1989}. However, Congress is a close-knit and social human institution. As such, we expect to observe tendencies toward intra-group homophily, in which members of a group display a preference for associating with other members of the same group.  This tendency, often described as ``birds of a feather flock together,'' is prevalent across social networks in a variety of contexts, from friendship groups to schools and businesses \citep{Goodreau:2008, Mollica:2003, Ruef:2003} and most commonly manifests as association with others of the same race or ethnicity \citep{McPherson:2001}.  Intra-group homophily has the potential to pose a particular problem in the legislative arena as it may negatively affect the ability of minority members to build support for their legislation.  If members of Congress draw support predominantly from colleagues of the same race, ethnicity, or gender, then minority members are disadvantaged first by their small group size and then by the tendency for the white, male majority to preferentially support other members of the majority.  This has serious implications for how well minority populations are represented even in the presence of descriptive representation.  

In this paper, we examine the degree to which members of demographic minorities are at a comparative disadvantage in the legislative process as a result of their minority status.  We find a strong tendency for intra-group homophily among members, yet the resulting disadvantage is mitigated by the electoral pressures that all members face.  Reelection-minded legislators cannot safely ignore sizable populations in their constituencies if they wish to maintain their electoral security.  Using the House cosponsorship network, we show that members display a strong bias towards their own race or ethnicity in their cosponsorship decisions.  At the same time, members who represent racially diverse districts are more likely to support colleagues of the same race as their constituents.  We also show that women do not face the same disadvantages as Black and Latino members, which we argue is the result of both the need to appeal to women constituents and an overall tendency against gender-based segregation.  As a result of these dynamics, Black and Latino members may struggle to build support for their legislation outside of their own in-group unless they appeal to colleagues who represent sizable minority populations.

Congress is the most professionalized legislature in the United States \citep{Squire:1992}.  The members who have been elected to Congress are, for the most part, professional politicians who have devoted substantial portions of their lives to successfully seeking election and rising through a series of democratically elected offices \citep{Berkman:1994}.  If these members are prone to favoring the legislation of their own in-group absent electoral pressures that compel them otherwise, then this suggests that de facto segregation not only imbrues relationships within the citizen population, but also between elected officials at all levels of government.  Yet as the population of the country becomes increasingly diverse, our research suggests that minorities will see increased support for their interests, provided that electoral districts are representative.  This is not to say that the election of minority members is not important, rather that both  descriptive representation of minority populations and increased diversity in electoral districts are necessary for leveling the playing field between minority legislators and their majority counterparts.

\section*{Representation of Minority Interests}
The degree to which minority members are disadvantaged is shaped by both internal and external dynamics of legislative bodies.  Internally, there is a wealth of evidence suggesting that minority legislators have a different experience than their white, male counterparts.  Black members are more likely to report experiencing discrimination \citep{Button:1996}, women of color report being marginalized in Congress \citep{Hawkesworth:2003}, and studies of women elected officials find that they tend to experience discrimination in the legislature \citep{Thomas:1994}, especially when it comes to seeking leadership positions \citep{Dodson:1991}.  Latinos and women have lower bill passage rates than their non-Latino and male counter parts \citep{Bratton:2006, Kathlene:1995} and peers rate black legislators as less effective than non-black legislators \citep{Haynie:2002}.  

At the same time, it is well established that members of Congress are motivated by a desire for reelection \citep{Mayhew:1974} and to win elections they must cultivate support among their constituents.  One way in which members may build support is through position taking, whether expressed through a member's voting record or simply by making a statement of support for a given policy.  Studies of representation have found that legislators generally try to remain in step with their constituents, from the seminal work of \citet{Miller:1963} to more recent work which finds that legislators are more likely to vote in line with their district if they are aware of constituent opinion \citep{Butler:2011}, and members who are ideologically aligned with their districts are more likely to make public statements of their positions \citep{Grimmer:2013}.  Members are also responsive to constituency opinion in their cosponsorship decisions \citep{Highton:2005} and cosponsorship is explicitly described by members and their staff as a valued form of position taking \citep{Koger:2003}.

In this study, we examine how this electoral connection between members and their constituents interacts with the tendency towards bias in the legislature to affect minority members in Congress.  The degree to which members support colleagues of a different race, ethnicity, or gender is the result of both a natural tendency towards intra-group homophily and the electoral pressures that they face.  For members who represent districts comprised predominantly of their own race or ethnic group, they have little incentive support colleagues of a different race or ethnicity.  Patterns of assortative mixing dominate; white and minority members alike display a bias in favor of supporting colleagues of their own race over members of a different group.  This form of race-based de facto segregation characterizes our neighborhoods \citep{Massey:1993}, schools \citep{Mollica:2003}, corporations \citep{Ruef:2003}, and social groups \citep{Louch:2000}. Thus, there is little reason to believe members of Congress are not prone to the same tendencies. Within the context of an institution as powerful and dominated by white males as the United States Congress, this tendency towards de facto segregation can put minority members at a particular disadvantage when they must secure support from their colleagues to advance their own legislative agenda.

Members representing racially diverse districts, however, cannot afford to neglect the interests of those constituents if they want to be reelected.  For a member representing a district with a sizable Black or Latino population, supporting legislation sponsored by Black or Latino members is an easy, low-cost way for a member to take a position and show support for that community.  Minority members are more likely to introduce legislation addressing minority issues \citep{Baker:2005, Rouse:2013}, Black voters tend to view Black representatives more positively than their white counterparts \citep{Tate:2003, Bowen:2014}, and Latino voters who are represented by co-ethnics report higher trust in government \citep{Ramirez:2012} and are more likely to correctly recall the race and party of their representative \citep{Bowen:2014}.  Supporting a Black or Latino colleague by signing onto the bills they introduce can be used by a member of a different race or ethnicity to send a signal to those constituents that (s)he is supportive of their interests and build electoral support among minorities.\footnote{We do not claim that constituents are closely following the bills that their members of Congress choose to cosponsor.  Rather, cosponsoring minority-sponsored legislation is a deliberate signal on the part of a member that (s)he can then make a point to emphasize when (s)he is reaching out to his or her minority constituents.}

While we expect to see clear patterns of assortative mixing and district responsiveness by race, the gender dynamics are somewhat more complicated.  Although women report discrimination in the legislature, studies on their effectiveness as legislators are less bleak than those for Blacks and Latinos.  Women are more successful in delivering federal spending to their district \citep{Anzia:2011}, and at passing legislation when in the minority party \citep{Volden:2013}.  Gender-based segregation is not as prevalent as race-based segregation.  Whereas someone may go their entire childhood without developing a meaningful relationship with someone outside of their race, males often develop relationships with female guardians and siblings, female schoolteachers and peers, and with females in their neighborhood or place of worship.  Therefore there is a certain familiarity that develops between men and women that may not develop between those of different races and ethnicities.  Furthermore, male members of Congress uniformly represent districts that are comprised of approximately 50\% women.  If any tendency towards assortative mixing is mitigated by electoral responsiveness, as we argue that it is, all members must concern themselves with their female constituents to some degree.  A male member who wishes to appear ``friendly'' to women, or supportive of women's issues has a strong incentive to support his female counterparts.

Therefore, we expect to observe three distinct patterns in the cosponsorship network.  First, Members of Congress will be less likely to cosponsor legislation introduced by members of another race or ethnicity than their own.  Second, Men will support legislation introduced by women at rates equal to bills sponsored by their own gender.  Third, as the proportion of a minority population in their district increases, the pressures of representation will push members to cosponsor more bills introduced by colleagues of that race or ethnicity.  The result is that racial and ethnic minority members remain at a disadvantage as a result of bias in cosponsorship patterns, but as constituencies become more diverse, that bias is mitigated.

\section*{Cosponsorship as a Show of Support}
The few studies on how minority constituency size translates into support for minority members of Congress have focused predominantly on Black populations and support for civil rights legislation.  Democrats with significant African American constituencies were more likely to vote for final passage of the Voting Rights Act \citep{Black:1978} and the 1990 Civil Rights Act \citep{Hutchings:1998}.  However, support for Black issues among members representing minority constituencies is inconsistent. On the one hand, research suggests that party matters more than race when it comes to voting on legislation, meaning that white Democrats represent the policy interests of African Americans \citep{Swain:1993, Grose:2011}. Other research on minority representation has found that increased minority population translates into decreased support for conservative legislation \citep{Combs:1984}, and incumbent Democrats became more supportive of Black issues when their district's Black population increased through redistricting \citep{Overby:1996}. On the other hand, research suggests that white Democrats in the South become responsive to black interests when it comes to roll call voting behavior once the group comprises forty percent of the population in a district \citep{Lublin:1997}, and even then their responsiveness is less certain when looking at lower profile issues \citep{Hutchings:2004}. The relationship between Latino population size and legislative responsiveness is complex as well. \citep{Hero:1995} show that Latino population has an indirect effect on whether the policy interests of the group are represented; the scholars show that in districts where Latinos comprise at least five percent of the population that Democrats are more likely to get elected, and in turn Democrats are more likely than Republicans to vote in a way that advocates for Latino interests. Later research shows that depending on the size of the Latino population, members of Congress become more or less responsive to Latino interests. In districts where Latinos are between forty and fifty percent of the population, members of Congress become less responsive to their interests, and yet once the group comprises fifty percent or more of a district's population, then members of Congress are responsiveness to Latino interests (through their roll-call voting behavior) \citep{Griffin:2007}.

Voting for a bill is only one of the ways a member can support a colleague and opportunities for support on the floor are limited to those bills that have survived an extensive winnowing process \citep{Krutz:2005}.  Cosponsorship, on the other hand, is both an important signaling device within the legislature and the only way to officially associate with a piece of legislation outside of casting a roll-call vote in favor of it.  Since most legislation dies in committee, and cosponsors can sign on at any point in the legislative process, cosponsorship is used to officially associate with legislation that will never make it to the floor.  Members cosponsor legislation both to send a signal within Congress that they support a bill \citep{Kessler:1996} and to show their support to constituents and other outside interests that can support their reelection \citep{Koger:2003}.  Although some have suggested that cosponsorship is merely ``cheap talk,'' members are clearly selective about the legislation they cosponsor, signing onto an average of 3.9\% of introduced bills from 1980 to 2004 \citep{Fowler:2006}.

In recent years scholars have developed several approaches to using observable legislative behavior to proxy networks of support and collaboration between legislators.\footnote{These include networks of co-voting \citep{ringe2013}, committee co-membership \citep{porter2005}, caucus co-membership \citep{ringe2013}, press event collaboration \citep{desmarais2015}, similarity in campaign contributions \citep{desmarais2015b}.} The most common approach to measuring collaboration between legislators has been through the use of cosponsorship data \citep{kirkland2014}. The cosponsorship stage of the legislative process also has considerable implications for legislative success \citep{Browne:1985, Crisp:2004,tam2010,kirkland2011}.  Members consider the list of cosponsors already on a bill when deciding whether to sign on \citep{Kessler:1996}, the number of cosponsors is a strong positive predictor of legislative success at the committee stage \citep{Wilson:1997}, and the connectedness of a legislator is positively associated with the success of amendments they introduce \citep{Fowler:2006}.  If minority members are unable to build widespread support for their legislation outside of their own in-group at this earliest stage of the legislative process, they may be discouraged from actively sponsoring legislation.

Several studies have examined cosponsorship patterns within minority groups in Congress.  \citet{Rocca:2008} find that Black and Latino members of Congress tend to cosponsor fewer bills than other members.  Numerous studies reveal that compared to men, women are more likely to cosponsor women's interest legislation \citep{Balla:2000, Wolbrecht:2000, Swers:2002, Swers:2005}.  Yet we know significantly less about how race, ethnicity, and gender influence the legislator to legislator component of cosponsorship. Research suggests that because of Congressional Black Caucus blacks may be inclined to support one another's legislation \citep{Canon:1999}, but this evidence is based on a quote and even then it is not clear that cosponsorship was being described. \footnote{The quote is from p.152 and as follows: "We have gotten good support from the CBC on our bills. The CBC is adept at understanding the political process and good at working together."} With this study, we study the effect of race, ethnicity, and gender both within and across minority groups and incorporate district demographics to provide a complete picture of race and gender based patterns of support in Congress.

\section*{Research Design}

To analyze the degree to which minority members are supported by their colleagues, we apply inferential methods of longitudinal network analysis to the US House cosponsorship network (HCN) for the 97th - 108th Congresses.\footnote{The corresponding years are 1981-2004.}  The cosponsorship process is fundamentally a \textit{relational} process because the act of cosponsoring another member's legislation - whether it implies a show of support for that legislator or for their legislation - represents a collaboration between the legislators.  It is commonly observed that the tally of cosponsors for any bill results from an explicitly interactive process between sponsor and cosponsors. Bill sponsors solicit support from their colleagues, who in-turn respond if they are willing to sign onto the legislation \citep{Fenno:1989, Koger:2003}.  Therefore, we follow \cite{Fowler:2006} in arguing that the set of complex relations between legislators is best represented as a network where each legislator is a node and each relationship is an edge.  

We use the House cosponsorship data from \citet{Fowler:2006, Fowler:2006b} and a temporal exponential random graph model (TERGM) to estimate the probability of cosponsorship from one member to another.\footnote{We briefly review the TERGM here.  For a more in depth discussion of the model, please see the supplemental appendix.}  The theoretical approach to the data that motivates the TERGM is that a network is not reducible to independent edges or subsets thereof, which makes it uniquely appropriate to handle a relational process such as cosponsorship.  Instead of having $2\binom{N}{2}$ observations in the sample, our sample contains twelve network-valued multivariate observations with each network capturing the cosponsorship relations in a two-year Congress.  Furthermore, the TERGM allows us to model both exogenous and endogenous network effects rather than assuming that Representative $i$'s decision to cosponsor Representative $j$'s bill is independent of other cosponsorship decisions \citep{Desmarais:2012}.  We can account for not only the attributes that make a member more or less likely to cosponsor legislation as traditional studies have done, but also the commonalities (or lack thereof) between cosponsor and bill sponsor, and the other relationships between members.

Whereas the exponential random graph model (ERGM) takes a single network as the measure of interest, the TERGM is an extension for longitudinal networks observed in $t$ discrete time periods \citep{Leifeld:2015}.  Each network $N$ is an adjacency matrix that records the number of times Representative $i$ cosponsored a bill introduced by Representative $j$ in Congress $t$.  However, the basic TERGM assumes ties between nodes to be binary and therefore, we recode the adjacency matrix so that observation $(i,j)$ is a binary indicator of whether Representative $i$ cosponsored more than one bill introduced by Representative $j$ in Congress $t$.  Because we are interested in support for minority members, we threshold the network at $>1$.  This eliminates the weakest connections between members so that a tie from Representative $i$ to Representative $j$ is a more meaningful indicator of support than cosponsoring a single bill would be.\footnote{Different thresholding specifications were tested and are discussed in the supplemental appendix. 
}  

The degree to which members support their colleagues through cosponsorship is then modeled as a function of both endogenous and exogenous dependencies, $\mathbf{h}(N)$.  Our specification allows us to estimate the effect of race and gender-related covariates on the probability of tie-formation from one member to another, while also accounting for network effects such as reciprocity, popularity, and triadic closure.  The probability of observing the network at some discrete period of observation $(N^t)$ is:
\begin{align*}
P(N^t | N^{t-K}, ..., N^{t-1}, \mathbf{\theta}) = \frac{exp(\mathbf{\theta}^T \mathbf{h}(N^t, N^{t-1}, ..., N^{t-K}))}{c(\mathbf{\theta}, N^{t-K}, ..., N^{t-1})}
\end{align*}

The model is estimated using maximum pseudolikelihood (MPLE) with bootstrapped confidence intervals, as developed by \citet{Desmarais:2012} and implemented in the xergm package \citep{Leifeld:2015}.  This approach deals with the computational intractability of the normalizing constant, $c(\mathbf{\theta}, N^{t-K}, ..., N^{t-1})$ by using a hill-climbing algorithm to find the maximum likelihood of the product over the conditional probability of each element given the rest of the network.  Because MPLE underestimates the variance of its estimates, a bootstrapped sample of MPLEs is used to calculate consistent confidence intervals for the parameter estimates.  The result is a set of model coefficients and corresponding confidence intervals that give us the likelihood of tie formation between nodes in the network, which in this case is an indicator of support from one member to another.

\section*{Determinants of Cosponsorship}
The decision to cosponsor legislation can best be understood as the result of three broad considerations: the cosponsoring member's own goals and ideals, the relationship between the cosponsor and bill sponsor, and the actions of colleagues in Congress.  Much of the previous empirical research on Congressional cosponsorship has modeled the aggregate tendency of legislators to cosponsor, irrespective of the sponsor and characteristics thereof \citep{Campbell:1982, Wilson:1997, Koger:2003}.  In network terms, this is referred to as a node's tendency to send ties, or out-degree covariates.  This work has produced consistent evidence that senior and electorally secure members cosponsor fewer bills, and that liberals and ideological extremists cosponsor more legislation \citep{Rocca:2008, Koger:2003, Campbell:1982, Kessler:1996}.  The relational component of cosponsorship has been less studied, although there is evidence that members with a history of supporting each other are less likely to renege on their cosponsorship pledges \citep{Bernhard:2013} and members of state legislatures are more likely to cosponsor bills from colleagues who are closer in ideological distance and represent neighboring districts, as well as those of the same race, ethnicity or gender \citep{Bratton:2011}

The network approach allows us to account for all three considerations through the covariates included in our model, which can be classified into three corresponding categories: out-degree covariates, dyadic covariates, and endogenous network effects.  Within this framework we are able to test our central hypotheses that the House cosponsorship network is characterized by race and gender based assortative mixing and that members are responsive to the racial composition of their district, while also demonstrating the necessity of a network approach to account for the interdependencies among members.

\begin{figure}
\centering
\includegraphics[scale = .45]{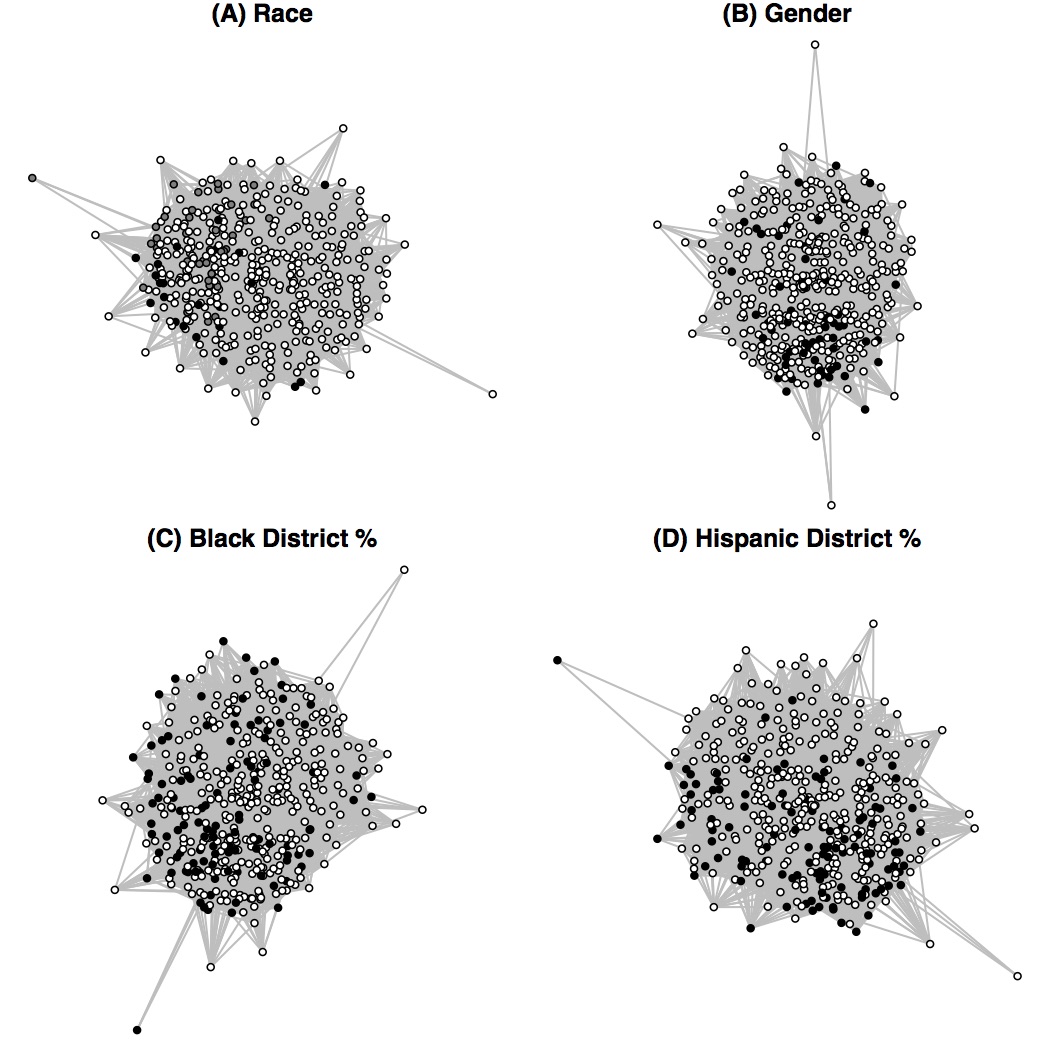}
\cption{\doublespacing Cosponsorship by Attribute in the 108th Congress}
\label{network}
\end{figure}

Visual representations of the cosponsorship support network from the 108th Congress, displayed in Figure \ref{network}, reveal the tendency for members to cluster together by race, gender, and district demographics.  Graph A colors the nodes by race, with white members in white, Black members in black, and Latino members in grey.\footnote{Asian, Pacific Islander, and Native American members are included with the white members as they do not comprise a large enough population in the House to estimate an effect over the time period studied here.} Most of the 38 Black and 24 Latino members of the 108th Congress are clearly clustered together on the left side of the graph, suggesting a strong tendency for these members to support each other.  Graph B plots the same network, this time highlighting the 60 women members in black.  Again, we observe a tendency for the women members to cluster together, although not as tightly as the racial and ethnic minorities do.  Graphs C and D show the network with the black nodes indicating members who represent districts in the upper quartile of Black and Hispanic constituencies (respectively).  For Black constituencies, this means a Black population greater than 14.8\% of the district and for Hispanic constituencies, this means a Hispanic population greater than 15.6\% of the district.  The members representing these districts with sizable minority populations also display a tendency to cluster together, although again, it is not as dense as the cluster formed by racial and ethnic minority members.

In addition to displaying race and gender based clustering in the cosponsorship network, the network graphs in Figure \ref{network} clearly show the underrepresentation of minorities and women in the House of Representatives.  If minority members are unable to attract cosponsors from the majority party, is that the result of discriminatory practices or simple numbers?  Baseline assortative mixing refers to the degree of assortative mixing which can be explained by the opportunity structure.  For example, if 10\% of the Congress were ethnic minorities, we would not expect an unbiased member of Congress to cosponsor half minority and half majority group legislation.  We would expect the cosponsorships of the unbiased member to be assortative 90\% of the time and disassortative 10\% of the time.  \textit{Preferential assortative mixing} on the other hand is assortative mixing beyond that which can be explained by the opportunity structure.  For example, if a member of Congress under the same circumstances as described above, cosponsored assortatively 98\% of the time, that would indicate a strong preference to work within one's ethnic group beyond the fact that Congress is mostly white. 

\begin{figure}
\centering
\includegraphics[scale = .75]{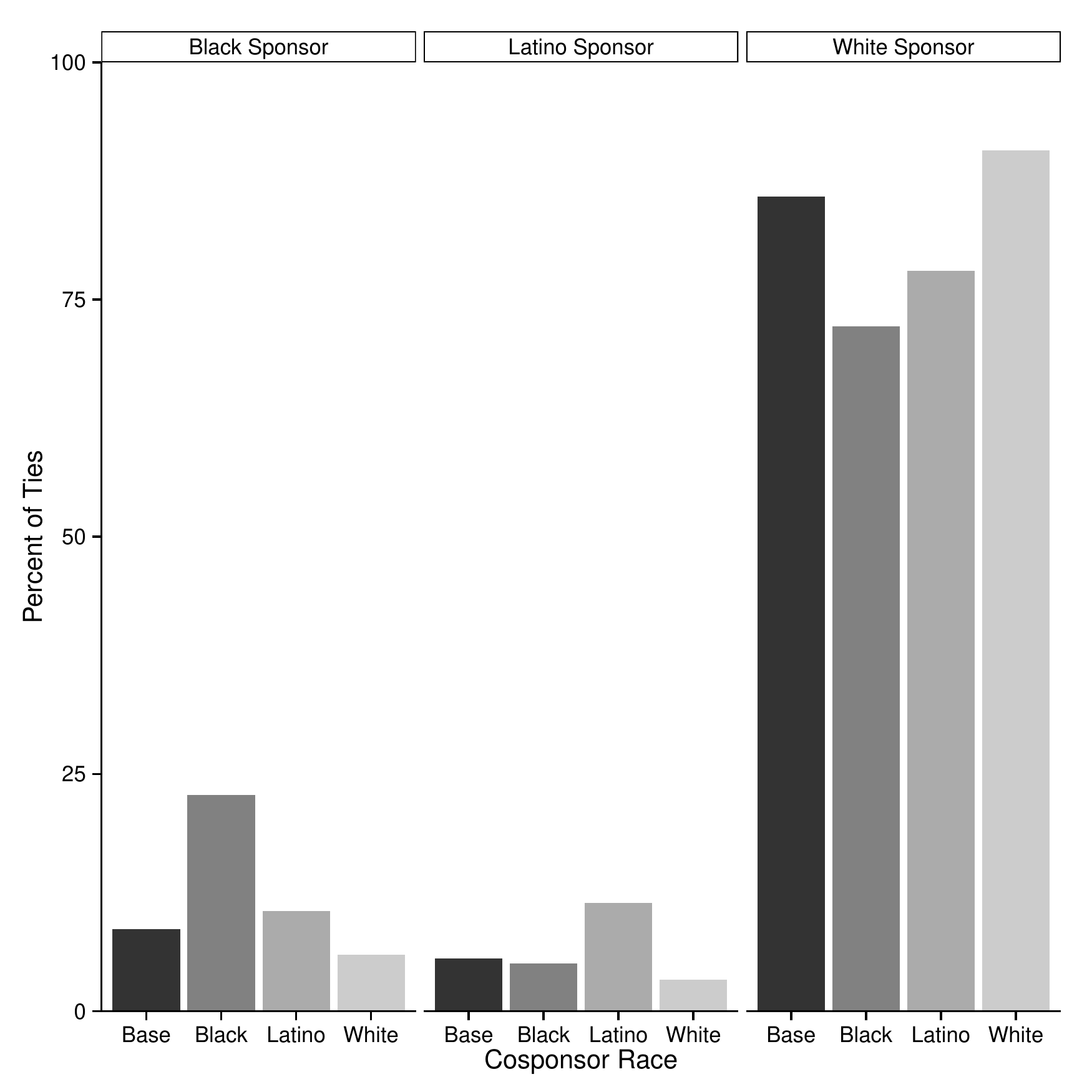}
\cption{\doublespacing Percent of Ties by Cosponsor Race Compared to Unbiased Baseline}
\label{racebar}
\end{figure}

To examine whether the House cosponsorship network shows baseline or preferential assortative mixing, Figure \ref{racebar} show the proportion of ties from each race to all other races, compared to the unbiased baseline.  In the cosponsorship network for the 108th Congress, there were 377 nodes coded as white\footnote{Which also includes four Asian/Pacific Islander and 3 Native American members}, 38 are Black, and 24 are Latino.  Therefore, for an unbiased member of Congress of any race, we should expect 85.88\% of their ties to be to white members, while 8.67\% are to Black members, and 5.55\% are to Latino members.  Instead, we see that 90.71\% of ties from white members are to other white members, 5.96\% are to Black members, and 3.32\% are to Latino members.  Black and Latino members display a similar tendency towards preferential assortative mixing.  For Black members, 22.81\% of their ties are to other Black members, 72.18\% are to white members, and 5.02\% are to Latino members.  For Latino members, 11.43\% of their ties are to other Latino members, 78.0\% are to white members, and 10.57\% are to Black members.  All three groups display a preference for supporting members of their own race, above what can be explained by the opportunity structure.

\begin{figure}
\centering
\includegraphics[scale = .75]{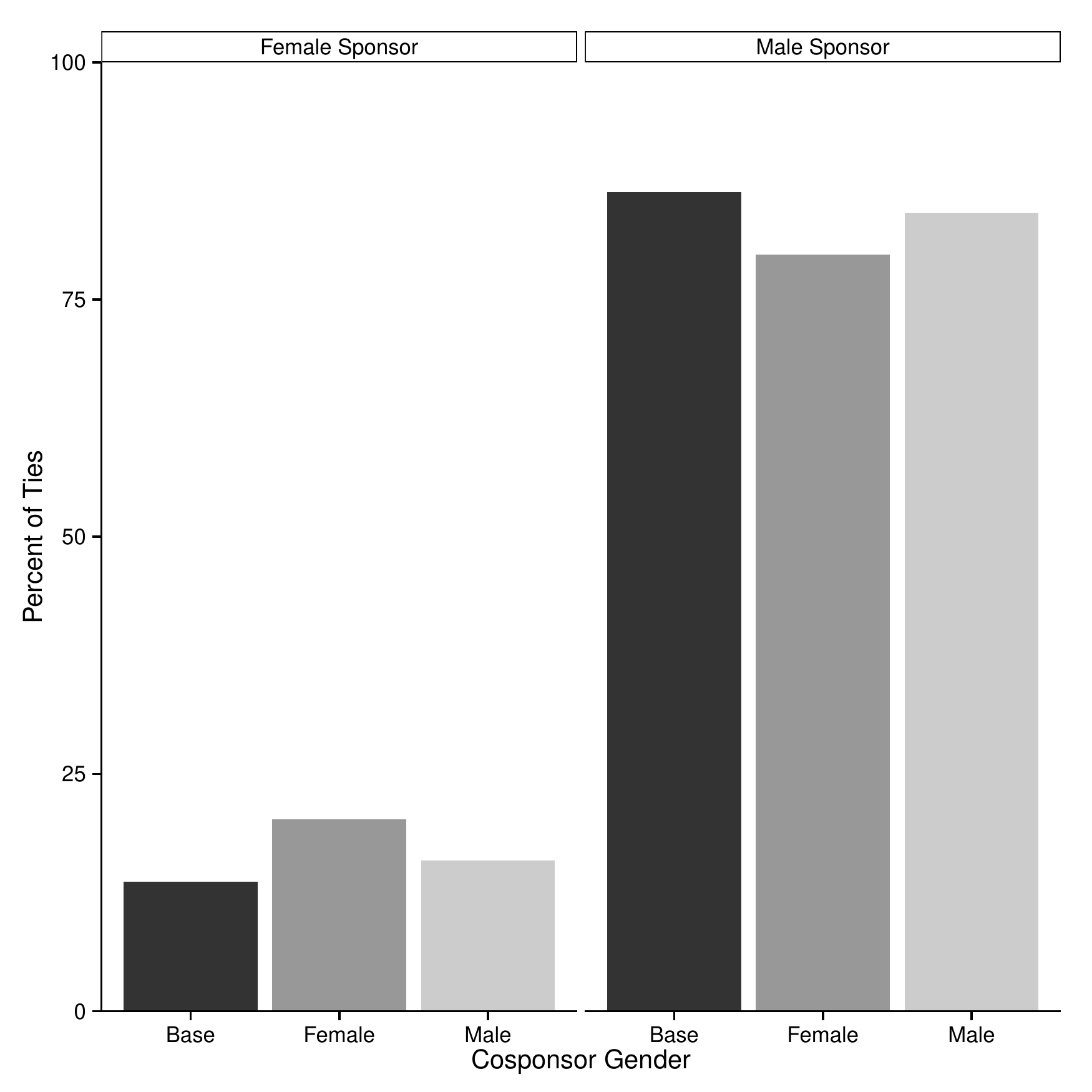}
\cption{\doublespacing Percent of Ties by Cosponsor Gender Compared to Unbiased Baseline}
\label{genbar}
\end{figure}

When considering the effect of gender, we observe a different pattern.  Here, our baseline expectation is that an unbiased member of Congress will give 86.33\% of their support to male members and 13.67\% of their support to female members.  As shown in figure \ref{genbar}, male members are close to this baseline, with 84.13\% of their ties going to their male colleagues and 15.87\% of their ties going to female colleagues.  However, female members display a tendency for preferential assortative mixing, giving 79.77\% of their support to male colleagues and 20.23\% of their support to female colleagues.  These patterns of assortative mixing by race and gender are supported by our empirical analysis, which we consider next.

\subsection*{Network Effects}
The first subset of covariates we need to consider is that comprised by the structural network factors.  The literature on intra-chamber dynamics provides great guidance in this phase of our model specification.  Three phenomena that factor prominently in the literature on Congressional processes are informational influence and cue-taking \citep{Kingdon:1973, Matthews:1975, Krehbiel:1987, Sullivan:1993, Bianco:1997}, mutual exchange \citep{Mooney:1991, Caldeira:1993, Gilligan:1994, Alvarez:1997} and coalition-building \citep{Hammond:1983, Ferejohn:1987, Baron:1989, Wiseman:2004}.  Three endogenous network effects that parallel these tendencies are popularity, reciprocity, and transitivity, respectively.

\textit{Popularity} in a social network is measured with in-degree statistics, which account for the number of directed ties to a given node.  Here, this corresponds to the number of unique cosponsors that a member has across all of the legislation that (s)he has introduced, which we operationalize using the geometrically weighted in-degree statistic.  This accounts for the number of ties that a node attracts, while placing a geometrically decreasing weight on nodes with higher degrees to avoid model degeneracy issues .  Within our vector of endogenous and exogenous network statistics, $\mathbf{h}(N)$, the geometrically weighted in-degree statistic adds a term equal to:
\begin{align*}
h_\alpha^{(d)}(N) = \sum_{k=0}^{n-1}e^{-\alpha k} d_k(N) = \sum_{i=1}^n e^{-\alpha N_{i+}}
\end{align*}
where $d_k(N)$ is the number of nodes with in-degree $k$ and $\alpha$ controls the rate of geometric decrease in the weights \citep{Snijders:2006}.  As an endogenous network statistic, what the popularity measures captures is the propensity for tie-formation to nodes simply because they are popular with the other nodes in the network.  

In the congressional context, this can be conceptualized as Representative $i$ cosponsoring Representative $j$'s legislation because (s)he sees that Representatives $k$, $l$, $m$, and $n$ have already done so.  Because we have thresholded the cosponsorship network to only include ties when a member has cosponsored more than one bill introduced by his or her colleague, we are able to distinguish between popular legislation and popular members.  We consider Representative $i$ a \textit{supporter} of Representative $j$ if (s)he cosponsored two or more of the bills sponsored by Representative $j$. The members who are able to attract a broad base of supporters can therefore be seen as cue-givers within the legislature.  In the 108th Congress, the member with the highest in-degree was Representative Phil English (R-PA) who attracted 333 supporters, compared to the median member, who attracted 38.  While this may be partly attributed to him sponsoring 61 bills, which is considerably higher than average, our model allows us to account for the possibility that his support built cumulatively, independent of other effects such as number of bills sponsored, ideology, race, or gender.

We also include a geometrically weighted out-degree statistic, which accounts for the number of ties that a node sends.  This captures \textit{sociality} in the network, or the tendency of some members to cosponsor legislation more freely, such as Representative Martin Frost (D-TX) who supported 221 of his colleagues in the 108th Congress and 205 in the 107th.  Some members are simply more selective than others in their cosponsorship decisions, and the sociality measure allows us to account for that.

\begin{figure}
\centering
\includegraphics[scale = .75]{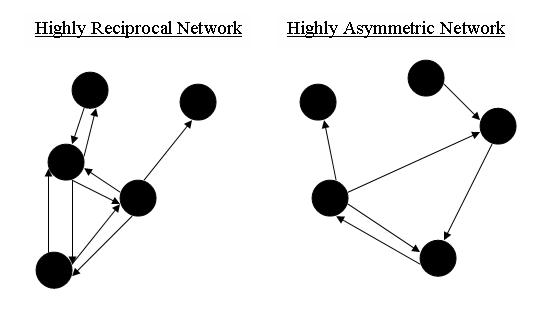}
\cption{\doublespacing Network Reciprocity}
\label{recip}
\end{figure}

The direct tendency for exchange in the network is operationalized as a measure of \textit{reciprocity}.  We add a statistic to the $\mathbf{h}(N)$ term that is equal to the number of pairs of actors $i$ and $j$ where $i \rightarrow j$ and $j \rightarrow i$.  This allows us to capture the degree to which members view supporting each other as a matter of mutual exchange.  A member who does not cosponsor the legislation of his or her colleagues may have trouble building support for his or her own bills, particularly if cosponsorship is seen as a favor to a colleague.  Again, as an endogenous network statistic, this allows us to consider patterns of reciprocity independent of other factors that might lead two members to support each other, such as ideological distance or serving on the same committee.  A positive coefficient on this indicates a tendency toward reciprocity in the network.  Figure \ref{recip} gives examples of small reciprocal and asymmetric networks.

\begin{figure}
\centering
\includegraphics[scale = .75]{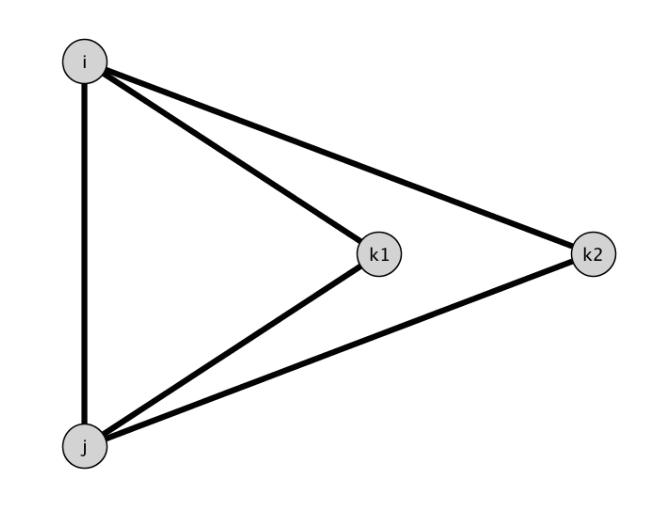}
\cption{\doublespacing Network Transitivity}
\label{triad}
\end{figure}

\textit{Transitivity} in a network provides a measure of clustering by accounting for the tendency of two nodes to share more than one partner.  In its simplest form of the transitive triad, we observe either $(i \rightarrow j, i \rightarrow k, k \rightarrow j)$ or $(i \rightarrow j, k \rightarrow j, k \rightarrow i)$.  We measure clustering using the geometrically weighted edgewise shared partner statistic (GWESP), which similarly to the geometrically weighted degree statistics, accounts for the tendency towards shared partners between nodes, while weighting the distribution to account for the higher likelihood that two nodes will have one shared partner than one hundred shared partners \citep{Hunter:2007, Wimmer:2010}.  The GWESP term adds a statistic equal to 
\begin{align*}
v(N; \phi_T) = e^{\phi T} \sum_{i=1}^{n-2} (1-(1-e ^{-\phi T})^i)EP_i(N)
\end{align*}
where $EP_i(N)$ is the number of edges between two nodes in the network that share exactly $i$ common partners \citep{Hunter:2008}.  Figure \ref{triad} illustrates the concept of edgewise shared partners by demonstrating a cluster in which nodes $i$ and $j$ have two edgewise shared partners, $k1$, and $k2$.

The equivalent relationships in Congress are an illustration of clustering that may be considered the result of either cue-taking or coalition building.  In a network prone to transitivity, if Representative $i$ supports both Representatives $j$ and $k$, then Representatives $j$ and $k$ should be more likely to support each other.  Representative $j$ may decide to support Representative $k$ because (s)he sees that Representative $i$ has done so and trusts his or her judgment, or Representative $i$ may forge a coalition with Representatives $j$ and $k$.  Untangling the mechanics of this relationship is beyond the scope of this paper, however we expect transitivity to figure prominently in the House cosponsorship network as a result of both cue-taking and coalition dynamics in Congress.

\subsection*{Legislator Effects} 
Many past studies have examined the influence of legislator characteristics on the tendency of legislators to cosponsor (out-degree).  In an analysis of the number of cosponsorships by legislator in the 95th House, \citet{Campbell:1982} found that majority party members cosponsor more than minority party members, senior legislators cosponsor less, and liberal and ideologically extreme legislators cosponsor more often.  \citet{Koger:2003} studies cosponsorship in the House over a period of ten Congresses and found that minority party members cosponsor more than majority party members, electorally vulnerable freshman cosponsor more, liberal members cosponsor more, and senior members cosponsor less.  \citet{Rocca:2008} examine cosponsorship in the 101st-108th Congresses and found that racial and ethnic minorities cosponsor less, women cosponsor more, senior members cosponsor less, and ideologically extreme legislators cosponsor more.  These findings are accounted for in our model as controls and to examine whether the dynamics of the cosponsorship process shift when accounting for the relational aspect of cosponsorship and setting a higher threshold for support.

\textit{Electoral margin} is coded as the cosponsor's share of the two-party vote, which is drawn from the \textit{Statistics of the General Election} compiled every two years by the Clerk of the House of Representatives.  We expect that members who have received a higher share of the two-party vote will cosponsor fewer bills and therefore support fewer of their colleagues.  As electorally secure members, they have less of a need to engage in the sort of position taking for constituents and interest groups that cosponsorship can represent.  Similarly, we expect that a cosponsor's \textit{Seniority}, which is coded as the number of Congresses they have served, will also have a negative effect on tie formation as most senior members already have well-established reputations and positions.  This has been one of the most consistent findings in the cosponsorship literature and we expect it to hold in our analysis.

Ideology and party have also been shown to play into the decision to cosponsor legislation and we include three controls here.  \textit{Ideology} is the cosponsor's 1st dimension DW-NOMINATE score, drawn from \citet{McCarty:1997}.\footnote{We test 2nd dimension DW-NOMINATE as well but exclude it from the final model as it has no effect on our results.}  \textit{Ideological extremity} is the absolute value of the difference between the cosponsor's 1st dimension DW-NOMINATE score and the chamber median.  \textit{Majority party} is a binary indicator for whether the cosponsor is a member of the majority party in that Congress.  We expect the ideology findings to be consistent with the existing literature that shows liberal and ideologically extreme legislators cosponsor more often than conservative and moderate legislators.  Liberal members cosponsoring more is consistent with an ideology that favors expanded government intervention, while ideologically extreme legislators are likely to view cosponsorship as a way to express their views on legislation that will not survive the winnowing process in the House.  For the same reason, we expect majority party members to cosponsor less than minority party members, whose favored policies are less likely to receive a floor vote.

Finally, we include controls for the racial composition of the cosponsor's district.  \textit{Black district percent} is the percent of the district population identifying as Black and \textit{Hispanic district percent} is the percent of the district population identifying as Hispanic.  These data are drawn from the Federal Election Project \citep{Lublin:2001} and the Congressional District Demographic and Political Data, 1972-1994 \citep{Lublin:1997}.  We do not have expectations on the direct effect of these variables and include them as a component part of the interaction terms discussed next. 

\subsection*{Relational Effects}
The primary covariates of interest in this study are those that apply to directed dyads of legislators.  Relational effects are captured with dyadic covariates, which are covariates that relate to a pair (or dyad) of nodes.  In the context of the House, whether two members are from the same state delegation is a dyadic covariate.  This variable would take the form of an NxN matrix where the ijth element is equal to one of Representatives $i$ and $j$ are from the same state.  Another example of a dyadic covariate is ideological distance.  Given the ideological positions of the $N$ members, a distance measure can be computed between the positions of Representatives $i$ and $j$.

To test our hypothesis that the House cosponsorship network is characterized by race and gender based assortative mixing, we include as dyadic edge covariates (1) a series of indicators for whether the tie from cosponsor to bill sponsor is one of nine possible permutations of race combinations, and (2) a series of indicators for whether the tie from cosponsor to bill sponsor is one of four possible permutations of gender combinations.  For example, the \textit{White $\rightarrow$ Black} covariate equals one if the tie is from a white cosponsor to a Black bill sponsor and zero otherwise.  This measure therefore captures the propensity of white members to support their Black colleagues.  Similarly, \textit{Men $\rightarrow$ Women} equals one if the tie is from a male cosponsor to a female bill sponsor and zero otherwise.  Every race and gender permutation is included in the model save the two largest of \textit{White $\rightarrow$ White} and \textit{Men $\rightarrow$ Men} which are the excluded reference categories.  Therefore, the results can be interpreted as the difference between the given mixing rate and the white $\rightarrow$ white rate for race and the men $\rightarrow$ men rate for gender.  If the confidence intervals for two mixing coefficients do not overlap, the rates are statistically distinct.  If the confidence intervals do not contain zero, then the rate is different from the intra-white rate or intra-male rate.  

We expect to see positive coefficients on the same race dyads of \textit{Black $\rightarrow$ Black} and \textit{Latino $\rightarrow$ Latino} and negative coefficients on the remaining six race mixing coefficients, in keeping with our hypothesis of race-based assortative mixing.  For the gender mixing coefficients, we expect to see little to no distinction between the rates at which men cosponsor women and other men and threfore \textit{Male $\rightarrow$ Female} should be statistically indistinguishable from our baseline excluded category of \textit{Male $\rightarrow$ Male}.

Our third hypothesis deals with the relationship between the racial composition of a cosponsor's district and the race of a bill sponsor.  We include four interaction terms to capture this relationship: \textit{Black District*White Sponsor}, \textit{Black District*Black Sponsor}, \textit{Latino District*White Sponsor}, and \textit{Latino District*Latino Sponsor}.  These interaction terms are created by multiplying the appropriate district racial composition variable for a cosponsor by the race of the bill sponsor.  The result is a variable that captures whether tie formation to a bill sponsor of a particular race is more likely as the percentage of that race in the district increases.  We expect \textit{Black District*Black Sponsor} and \textit{Latino District*Latino Sponsor} to be positive, with a corresponding decrease in the likelihood of cosponsoring bills introduced by white members.

We also include several variables that could confound results if omitted from the analysis.  To adjust for preference-based homophily we include as dyadic covariates the absolute difference between the 1st dimension DW-NOMINATE score for both legislators, with the expectation that legislators will be more likely to support those who are ideologically similar.  We also adjust for whether a pair of legislators are from the same party, or serve on the same committee, which we also expect to be positively associated with tie-formation.

Our network approach also allows us to adjust for sponsor characteristics that are likely to influence the support they receive from other members in the House.  First and foremost is the \textit{Bills Sponsored} by the member.  These data are drawn from the Congressional Bills Project \citep{Adler:1991} and are included in the model as an in-degree covariate as a count of the number of bills a member introduced in that Congress.  Obviously members who introduce more legislation will provide more opportunities for their colleagues to support them than members who introduce only a few bills.  Because of our interest in race effects, we also adjust for the number of race bills introduced, \textit{Race Bills Sponsored}, using the bill topic coding done by the Policy Agendas Project \citep{Baumgartner:2013}.  To account for the possibility that the number of race bills a member sponsors may have different effects in attracting supporters, depending on their race, we include two more interaction terms, \textit{Latino*Race Bills} and \textit{Black*Race Bills} which interact the race of the cosponsoring node with the number of bills introduced by the sponsoring node.

Finally, we include a series of network-level controls: \textit{Congress} and its associated polynomials, \textit{Congress$^2$} and \textit{Congress$^3$} adjust for the variation in patterns of support from one Congress to the next.  \textit{Party Homophily} interacts whether two members are from the same party with the Congress in which they serve.  

\section*{Results} 
\begin{figure}
\centering
\includegraphics[scale = .75]{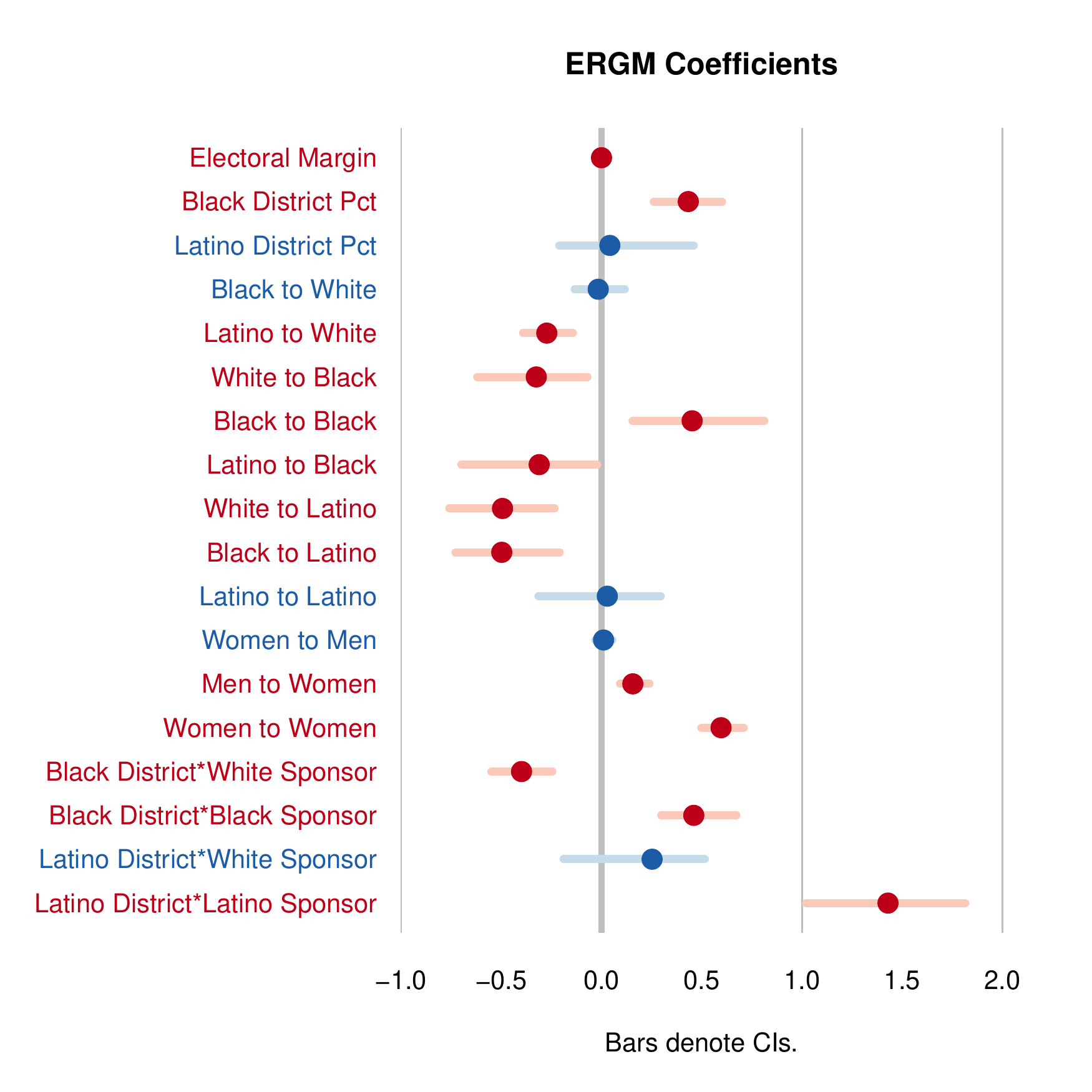}
\cption{\doublespacing Coefficients and 95\% confidence intervals for key independent variables}
\label{coefplot}
\end{figure}

We argue that racial and ethnic minority members of Congress are at an inherent disadvantage as a result of assortative mixing patterns, but that this disadvantage is mitigated by the electoral pressures that members representing diverse districts face, while women do not face a similar disadvantage.  The results of our TERGM estimation support our hypotheses and are presented in Table \ref{results}.  The coefficient estimates represent the change in the conditional log odds of tie formation in the network as a result of a unit change in the respective covariate or network statistic. The coefficients are presented with the 95\% confidence intervals calculated using 1,000 bootstrap iterations.  We also present a plot of our key coefficients of interest in Figure \ref{coefplot} for ease of interpretation.

\begin{table}
\begin{center}
\begin{tabular}{l r r r }
\hline
                        & Estimate & 2.5\% & 97.5\% \\
\hline
Edges                   & -3.4146  &-4.4151  &-2.5182    \\
Reciprocity                  & 0.7378 &0.6735 &0.7981       \\
Sociality               & -33.277 &-44.8033 &-24.6346 \\
Popularity               & -16.268 &-18.5485 &-13.8439 \\
Transitivity           & 0.6054 &0.5294 &0.6861      \\
Electoral Margin        & -0.0007 &-0.0013 &-0.0003    \\
Seniority         & -0.0310 &-0.0372 &-0.0243    \\
Ideology         & -0.4615 &-0.5557 &-0.3537    \\
Majority Party       & 0.0483 &-0.0700 &0.1618         \\
Ideological Extremity        & 1.1596 &0.9127 &1.4197       \\
Percent Black Population           & 0.4324 &0.2653 &0.5925       \\
Percent Hispanic Population           & 0.0409 & -0.2390& 0.4340          \\
Bills Sponsored      & 0.0296 & 0.0257& 0.0344      \\
Race Bills Sponsored          & 0.0734 & -0.0474& 0.1553        \\
Latino * Race Bills     & 0.0415 & -0.0385& 0.1270 \\
Black * Race Bills     & 0.0587 & -0.0463& 0.1653          \\
Same Committee    & 0.3907 & 0.3596& 0.4315       \\
Ideological Distance          & -2.0388 & -2.2251& -1.9029   \\
Same Party         & 0.0008 & -0.0902& 0.1071         \\
Black $\rightarrow$ White            & -0.0169 & -0.1342& 0.1108         \\
Latino $\rightarrow$ White            & -0.2737 & -0.3979& -0.1467    \\
White $\rightarrow$ Black            & -0.3262 & -0.6222& -0.0792    \\
Black $\rightarrow$ Black            & 0.4516 & 0.1527& 0.8091       \\
Latino $\rightarrow$ Black            & -0.3117 & -0.6998& -0.0216    \\
White $\rightarrow$ Latino            & -0.4946 & -0.7496& -0.2408    \\
Black $\rightarrow$ Latino            & -0.4984 & -0.7246& -0.2223    \\
Latino $\rightarrow$ Latino            & 0.0282 & -0.3196& 0.2751          \\
Women $\rightarrow$ Men          & 0.0096 & -0.0323& 0.0553          \\
Men $\rightarrow$ Women          & 0.1555 & 0.0947& 0.2376       \\
Women $\rightarrow$ Women          & 0.5960 & 0.4987& 0.7088       \\
Black District * White Sponsor      & -0.3997 & -0.5489& -0.2334    \\
Black District * Black Sponsor      & 0.4598 & 0.3114& 0.6918       \\
Latino District * White Sponsor      & 0.2518 & -0.1987& 0.5383          \\
Latino District * Latino Sponsor      & 1.4290 & 0.9790& 1.8145       \\
Congress   & 0.6581 & 0.1628& 1.2266       \\
Congress$^2$ & -0.1082 & -0.2243& -0.0338    \\
Congress$^3$ & 0.0051 & 0.0013& 0.0114       \\
Party Homophily  & -0.0229 & -0.0396& 0.0019        \\
\hline
\end{tabular}
\caption{TERGM estimates, 97th-108th congresses.  Coefficient estimates and 95\% bootstrap confidence intervals (1,000 bootstrap iterations) are given.}
\label{results}
\end{center}
\end{table}

Our first hypothesis is that members are less likely to cosponsor legislation introduced by members of another race or ethnicity than their own. This hypothesis is strongly supported for Black members and somewhat supported for Latino members.  We see that all but one of the heterophilous race combinations are negative and statistically distinct from zero, indicating that these pairings are less likely to occur than our baseline category of \textit{White $\rightarrow$ White}.  The sole exception is the relationship from Black to white members, which is indistinct from the relationship from white to white members.  Most relevant to our theory of minority members as being at a disadvantage are the coefficients of \textit{White $\rightarrow$ Black} and \textit{White $\rightarrow$ Latino}.  Here we find that white members are less likely to cosponsor bills introduced by minority members than their own race.  There is also a clear pattern of assortative mixing among Black members, who support their Black colleagues at rates greater than either White or Latino members do, while \textit{Latino $\rightarrow$ Latino} is indistinct from either white in-group support or support for Latino members from other races.  Therefore we conclude that white members are less likely to support minority members than their own race, Black members receive less support from their white and Latino colleagues than their own race, while there is no distinction between rates of support for Latino members across all three races.

Racial and ethnic minorities represent such a small proportion of the legislature that they must rely on support from their white colleagues to advance their agenda.  Our analysis here shows that minority members of Congress, particularly Black members, are at an even greater disadvantage than numbers alone would indicate.  The same racial assortative mixing patterns that we observe in other social networks from online communities to corporations are prevalent in the United States Congress and the result is de facto segregation in which members prefer to support colleagues of their own race to the detriment of minority members.

Our second hypothesis regards gender effects in the cosponsorship network.  We expect that men will support legislation introduced by women at rates equal to bill sponsored by their own gender.  Instead we find the somewhat surprising result that men are even more supportive of their women colleagues than they are of their own gender.  The effect is slight in magnitude, but \textit{Men $\rightarrow$ Women} is statistically distinct from \textit{Men $\rightarrow$ Men}.  On the other hand, assortative mixing is prevalent among women, who support each other at rates much higher than any of the other gender mixing coefficients.

We posit two explanations for why men are even more supportive of their women colleagues in Congress than we expected.  The first is that de facto segregation based on gender is far less prevalent than segregation based on race and ethnicity.  Whereas someone may spend their formative years without developing a meaningful relationship with someone outside of their own race, men frequently develop relationships with women in their families, schools, and neighborhoods.  Consequently, while it is feasible for someone to develop feelings of disattachment based on race and ethnicity, they are much less likely to do so based on gender as a result of the socialization that occurs between men and women.  As a result, while we observe women display a preference for their own gender that we argue is due to their status as minorities in the legislature, men do not display a similar preference for supporting their male colleagues.  The second explanation is that male support for female colleagues is the result of electoral pressures similar to the electoral-racial dynamics that we observe.  Whereas some members represent districts with larger minority populations than others, and as a result we are able to disentangle the effect of district composition by race, the proportion of women in Congressional districts is nearly constant across all 435 House districts.  Therefore, all men represent districts in which women comprise half of the population and they cannot afford to ignore this constituency.  However, as a result of the lack of variation in district gender composition, we are unable to conclusively demonstrate that male support for their female colleagues is the result of electoral pressures.

Finally, we consider our third hypothesis, that as the proportion of a minority population in the district increases, members will be more likely to support colleagues of that race or ethnicity.  Here we look at the four district composition-sponsor race interaction terms.  From Table \ref{results} we see that increasing the percentage of a minority population in a Representative's district has a significant effect on the probability that they support a colleague of that race and ethnicity.  Both \textit{Black District*Black Sponsor} and \textit{Latino District*Latino Sponsor} are positive and statistically significant.  They are also statistically distinct from the effect of district composition on support for white members in both cases.  As the proportion of Black members in a district increases, the member representing that district is significantly more likely to support a Black colleague than they are a white colleague.  The same pattern holds for members representing districts with sizable Latino populations, who are significantly more likely to support Latino colleagues than white members.

While the coefficient estimates show us the general tendency of members to support colleagues of a different race, gender or ethnicity, they tell us little about the substantive magnitude of these effects.  To better understand the effects of race, gender, and district composition in the cosponsorship network, we calculate the predicted probabilities of tie formation by sampling dyadic pairs and calculating the probability of tie formation between those two nodes \citep{Desmarais:2012b}.  Using this sampling procedure we are able to estimate the probability of tie formation between nodes of different race, ethnicity, or gender at each of the twelve time periods in our network.

\begin{figure}
\centering
\includegraphics[scale = .75]{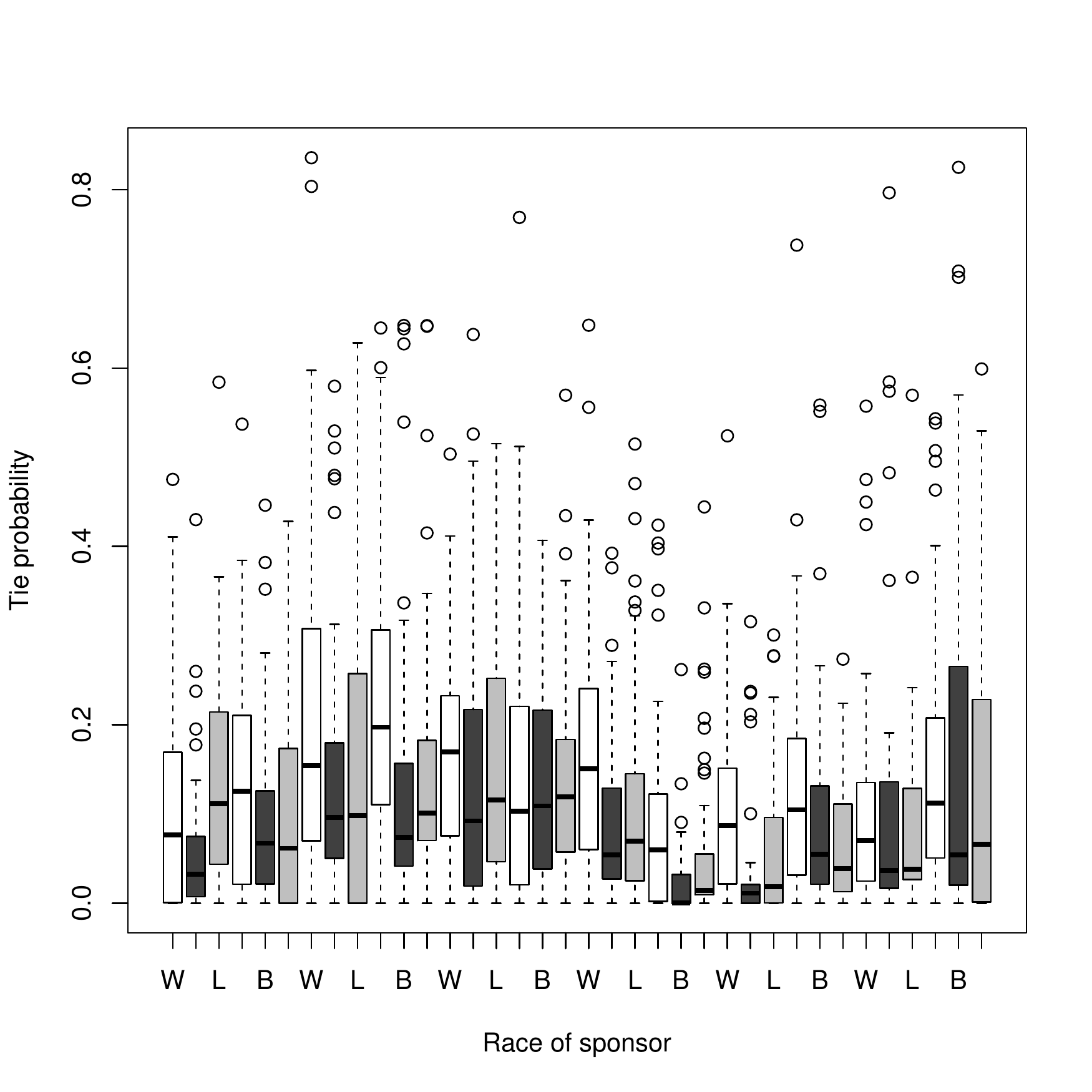}
\cption{\doublespacing Probability of tie formation from white cosponsors to white, Black, and Latino bill sponsors for the 97th through 108th Congresses}
\label{whiteout}
\end{figure}

Figure \ref{whiteout} shows that the given the rest of the network and the other terms included in our model, the median probability for ties from white cosponsors to white bill sponsors is consistently higher than the probability of ties from white cosponsors to Black or Latino bill sponsors, with the exception of the 97th and 102nd Congresses.  Aggregated over the whole time period, the median predicted probability of a tie from a white cosponsor to a white sponsor is 0.11, compared to a 0.05 probability of a tie from a white cosponsor to a Black sponsor, and a 0.07 probability of a tie from a white cosponsor to a Latino sponsor.  Again, we observe a clear tendency for white members to favor their white colleagues in their cosponsorship decisions.

The tendency for assortative mixing is even stronger among Black and Latino members, who display a clear preference for supporting colleagues of their same race or ethnicity.  For a Black cosponsor, the median probability of supporting a Black colleague is 0.60, compared to a 0.10 probability of supporting a white colleague, and a 0.24 probability of supporting a Latino colleague.  Latino bill cosponsors are similarly supportive of their same-race compatriots.  We observe the median probability of tie formation from a Latino cosponsor to a Latino bill sponsor to be 0.36, compared to 0.10 for white sponsors, and 0.16 for Black sponsors.  However this preference for supporting same-race colleagues among Black and Latino members is not enough to overcome the disadvantage they face in building support for their legislation as a result of their small population in the U.S. House.

\begin{figure}
\centering
\includegraphics[scale = .75]{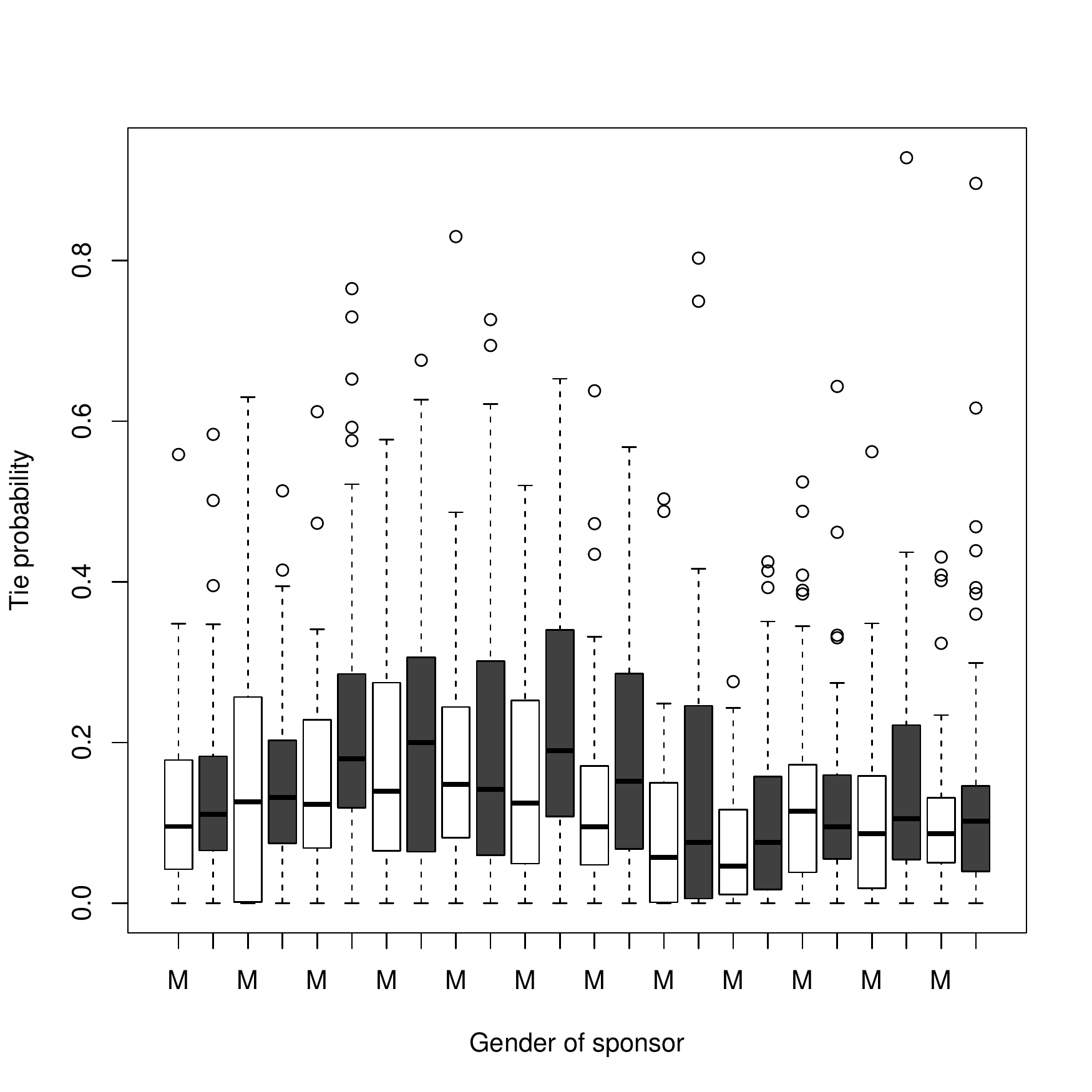}
\cption{\doublespacing Probability of tie formation from male cosponsors to male and female bill sponsors for the 97th through 108th Congresses}
\label{genties}
\end{figure}

We observe a different pattern when we consider the probability of tie formation by gender.  Women display a strong tendency for assortative mixing, with a 0.27 median probability of tie formation from one woman member of Congress to another.  Men, on the other hand, are slightly more likely to support their female counterparts than other men.  The median probability of tie formation from a male cosponsor to a male bill sponsor is 0.10, compared to a 0.12 probability of a male member supporting a woman.  Figure \ref{genties} shows the probability of tie formation by gender across all twelve Congresses. We see a consistent pattern of men supporting women at a slightly higher rate.  As expected, gender-based segregation is not as prevalent as race-based segregation in the U.S. House of Representatives.

\begin{figure}
\centering
\includegraphics[scale = .75]{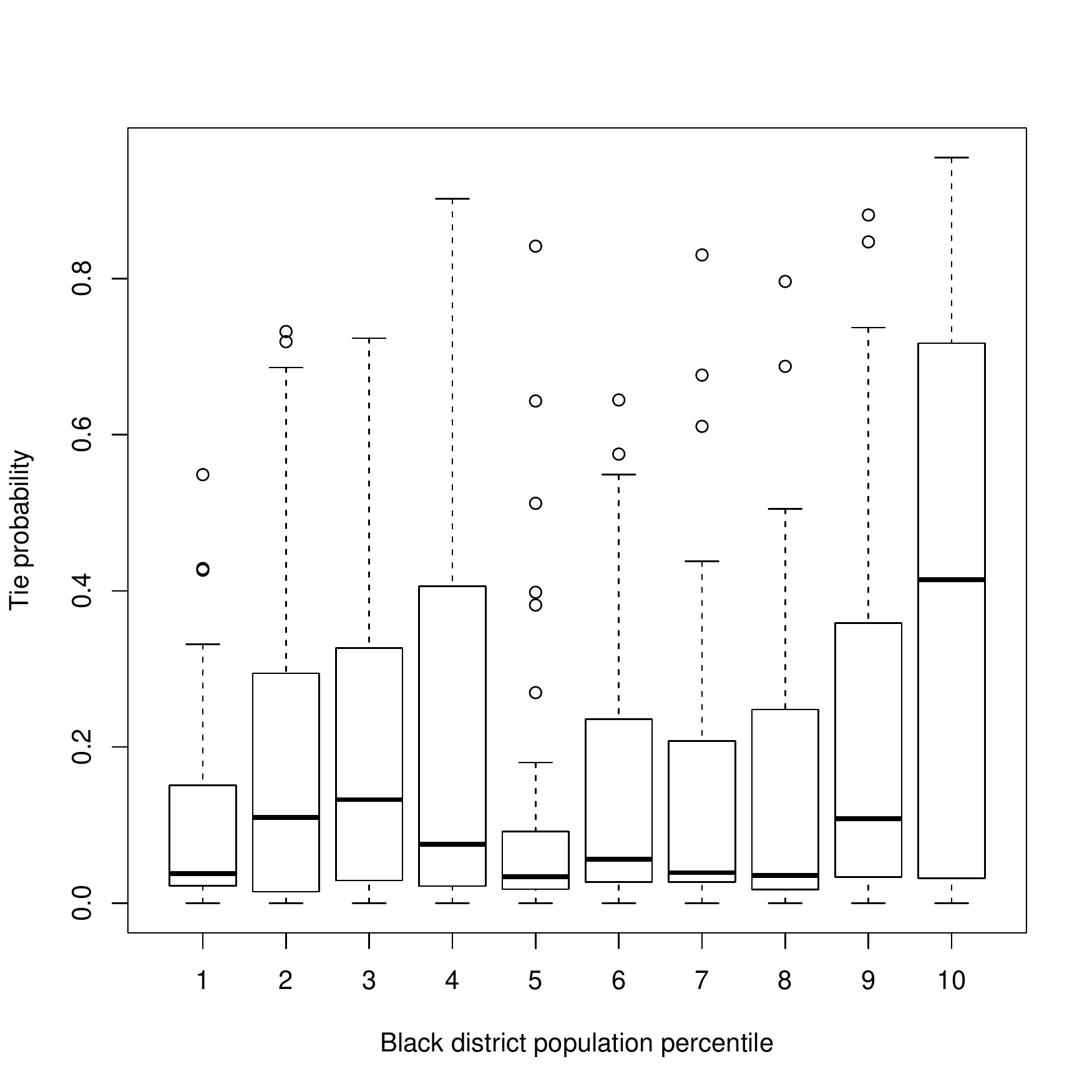}
\cption{\doublespacing Probability of tie formation from any race cosponsor to Black bill sponsors by district percent black population (in deciles) for the 108th Congress}
\label{bdpct}
\end{figure}

\begin{figure}
\centering
\includegraphics[scale = .75]{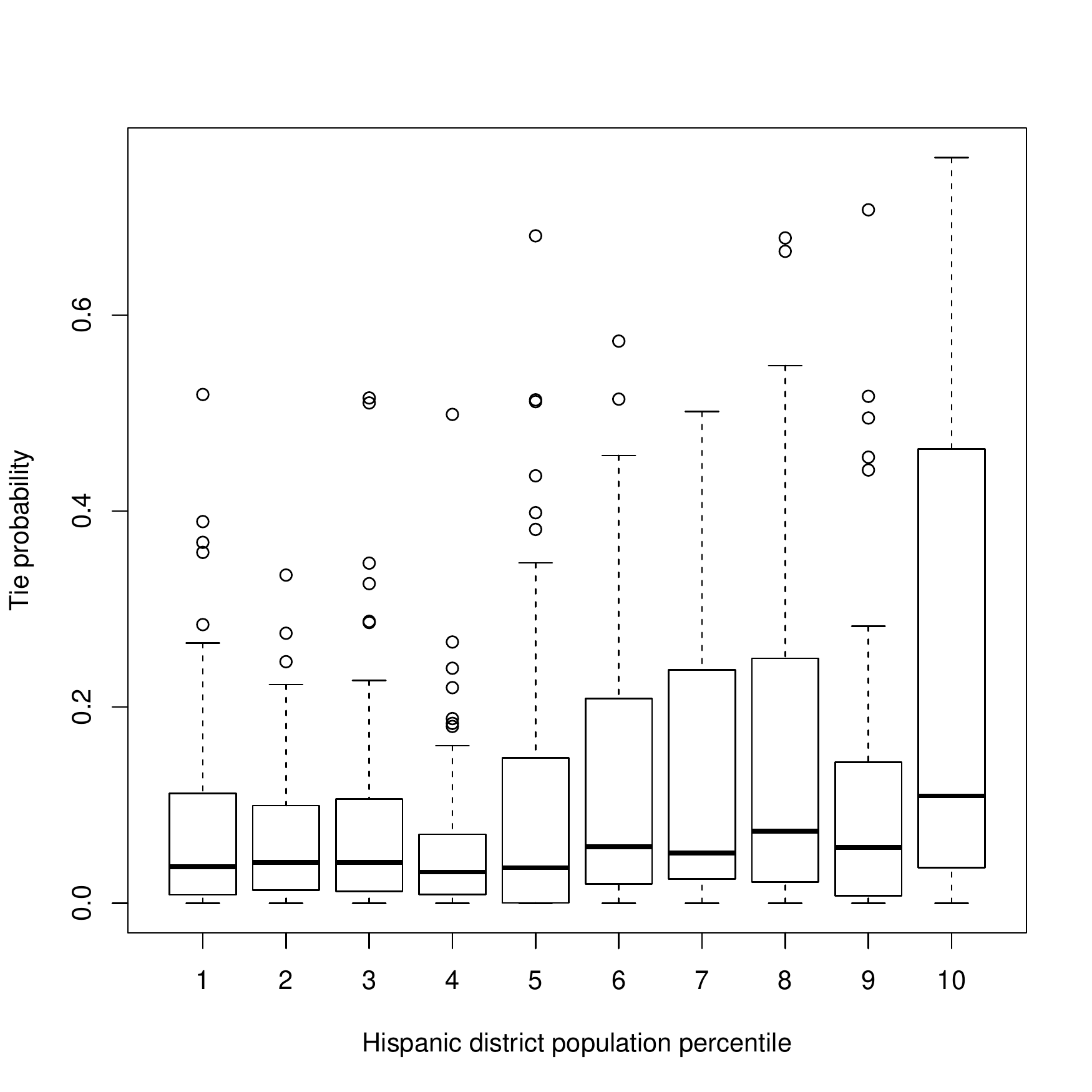}
\cption{\doublespacing Probability of tie formation from any race cosponsor to Latino bill sponsors by district percent hispanic population (in deciles) for the 108th Congress}
\label{hdpct}
\end{figure}

Finally, we consider the probability of support for Black and Latino members as the population of Black and Latino constituents increases in a member's district.  Here we focus on the 108th Congress in particular and calculate the predicted probabilities of tie formation from a cosponsor to a Black bill sponsor as the percentage of Black constituents increases as well as the predicted probabilities of tie formation from a cosponsor to a Latino bill sponsor as the percentage of Latino constituents increases.  Figures \ref{bdpct} and \ref{hdpct} show the predicted probabilities of ties to Black and Latino representatives (respectively) at each decile of Black and Latino district population.  The probability of tie formation to a Black bill sponsor remains relatively constant until the 90th percentile, which in the 108th Congress represents districts with a Black population of 32.66\%.  Here, the median probability of a member supporting a Black bill sponsor is 0.41, compared to 0.04 in districts with a Black population of 1.4\% or less.  When looking at the effect of Hispanic district percentage, we see a more consistent positive trend but also one that is smaller in magnitude.  The probability of tie formation to a Latino bill sponsor from a member representing a district that is less than 1.3\% Hispanic is 0.04, increasing to 0.11 for members representing districts that are over 34.54\% Hispanic.  As expected, members representing districts with sizable minority populations are much more likely to support colleagues of that race or ethnicity.

\section*{Conclusion}
The underrepresentation of minorities in the United States Congress affects not only the degree to which minorities are descriptively represented by their elected leaders, but also the effectiveness of minority leaders in the legislative process.  As we have shown, members of Congress display a strong tendency for assortative mixing in their cosponsorship decisions, preferring to support colleagues of the same race over and above what would be expected from an unbiased member in a disproportionately white and male legislature.  Black and Latino members are disadvantaged first by their small numbers in the House of Representatives and second by the tendency for the white majority to favor other white members.

Unlike racial and ethnic minorities, while women members are still at a disadvantage as a result of the disproportionately male membership of Congress, men do not display the same pattern of assortative mixing towards women as white members do towards Blacks and Latinos.  In fact, male members display a slight preference for supporting their female colleagues over their fellow men.  Assortative mixing by gender is both less prevalent than assortative mixing by race and is likely to result in less bias in constituency representation as members uniformly represent districts that are close to 50\% women.  When considered in tandem with the evidence of increased support for minority members as the size of a member's minority constituency increases, we conclude that electoral pressures can mitigate at least some of the disadvantages that minority members face. 

This research has implications for studies of redistricting. Arguably the most important factor that determines the presence of racial and ethnic minorities in office is the presence majority-minority districts comprised of the racial and ethnic groups in question \citep{Casellas:2011, Lublin:1997}. Several studies have considered how the race or ethnicity of a legislator affects the way constituents relate to that individual \citep{Tate:2003, Bowen:2014, Gay:2002}, and our study contributes to this rich literature, albeit in a different manner. Our findings suggest that aside from the quality of substantive representation blacks and Latinos receive from elected officials, another consequence of racial gerrymandering is that it alters the way legislators relate to one another. Thus, not only does racial gerrymandering influence the constituent-legislator relationship, but it also alters the legislator to legislator relationship. 

Minority members of Congress are disadvantaged on several fronts, from reports of outright discrimination to obtaining leadership positions and passing legislation.  With this paper, we add the ability of Black and Latino members to obtain support through the cosponsorship process to the list of ways that minorities struggle in Congress.  All members must rely on their colleagues to advance their legislative agendas, and this trend of race-based assortative mixing has particularly troubling implications for the effectiveness of minority legislators.

At the same time, we demonstrate a clear pattern of members responding to the racial composition of their district, which suggests that members are able to use cosponsorship of minority interest bills for their electoral benefits and that members who represent diverse districts are particularly sensitive to their minority constituencies.  As the minority population of the country continues to increase, we expect that whatever disadvantage minority members face will continue to be lessened as their colleagues represent increasingly diverse districts and use support for minority members and minority issues to shore up support for their own reelection.

\clearpage
\doublespacing
\bibliographystyle{apsr}
\bibliography{cospbib-1,cospbib}
\end{document}